\documentclass[aps,showpacs,twocolumn,prl]{revtex4-1}%

\usepackage{amsmath}
\usepackage{amsfonts}
\usepackage{amssymb}
\usepackage{times}
\usepackage{natbib}%
\providecommand{\U}[1]{\protect\rule{.1in}{.1in}}
\usepackage[dvips]{graphicx}
\DeclareGraphicsExtensions{.eps}
\begin{document}
\title{Beating the efficiency of both quantum and classical simulations with semiclassics}
\author{Cesare Mollica}
\author{Ji\v{r}\'{\i} Van\'{\i}\v{c}ek}
\email{jiri.vanicek@epfl.ch}
\affiliation{Laboratory of Theoretical Physical Chemistry, Institut des Sciences et
Ing\'{e}nierie Chimiques, Ecole Polytechnique F\'{e}d\'{e}rale de Lausanne,
Lausanne, Switzerland}
\date{\today }

\begin{abstract}
While rigorous quantum dynamical simulations of many-body systems are
extremely difficult (or impossible) due to the exponential scaling with
dimensionality, corresponding classical simulations completely ignore quantum
effects. Semiclassical methods are generally more efficient but less accurate
than quantum methods, and more accurate but less efficient than classical
methods. We find a remarkable exception to this rule by showing that a
semiclassical method can be both more accurate and faster than a classical
simulation. Specifically, we prove that for the semiclassical dephasing
representation the number of trajectories needed to simulate quantum fidelity
is independent of dimensionality and also that this semiclassical method is
even faster than the most efficient corresponding classical algorithm.
Analytical results are confirmed with simulations of quantum fidelity in up to
$100$ dimensions with $2^{1700}$-dimensional Hilbert space.

\end{abstract}
\maketitle


\emph{Introduction}. Correct description of many microscopic dynamical
phenomena, such as ultrafast time-resolved spectra or tunneling rate
constants, requires an accurate quantum (QM) simulation. While classical
(CL)\ molecular dynamics simulations are feasible for millions of atoms,
solution of the time-dependent Schr\"{o}dinger equation scales exponentially
with the number $D$ of degrees of freedom (DOF) and is feasible for only a few
continuous DOF. An apparently promising solution is provided by semiclassical
(SC) methods, which use CL trajectories, but attach to them phase information,
and thus can approximately describe interference and other QM effects
completely missed in CL simulations. Unfortunately, SC methods suffer from the
\textquotedblleft dynamical sign problem\textquotedblright\ due to the
addition of rapidly oscillating terms, resulting in the requirement of a huge
number of CL trajectories for convergence. Consequently, most SC methods are
much less efficient than CL simulations and in practice were used for at most
tens of DOF. Even though several techniques have explored this
issue~\cite{walton:1996,*kay:1994a,*sklarz:2004,*wang:1998,
*tatchen:2011,*tao:2011,*vanicek:2001,*vanicek:2003}, the challenge remains
open. Below we turn this challenge around by showing that in simulations of QM
fidelity (QF) \cite{gorin:2006,jacquod:2009}, a SC method called
\textquotedblleft dephasing representation\textquotedblright\ (DR) is not only
more accurate but, remarkably, also \emph{faster} than the most efficient
corresponding CL algorithm~\cite{mollica:2011}.

\emph{Quantum and classical fidelity.} QF was introduced by
Peres~\cite{peres:1984} to measure the stability of QM dynamics (QD). He
defined QF $F_{\text{QM}}(t)$ as the squared overlap at time $t$ of two QM
states, identical at $t=0$, but subsequently evolved with two different
Hamiltonians, $H_{0}$ and $H_{\epsilon}=H_{0}+\epsilon V$:
\begin{align}
F_{\text{QM}}(t)  &  :=\left\vert f_{\text{QM}}(t)\right\vert ^{2}%
,\label{QF}\\
f_{\text{QM}}(t)  &  :=\langle\psi\left\vert U_{\epsilon}^{-t}U_{0}%
^{t}\right\vert \psi\rangle, \label{fQM}%
\end{align}
where $f_{\text{QM}}(t)$ is the fidelity amplitude and $U_{\epsilon}^{t}%
:=\exp(-iH_{\epsilon}t/\hbar)$ the QM evolution operator. Rewriting Eq.
(\ref{fQM}) as $f_{\text{QM}}(t)=\langle\psi\left\vert U^{t}\right\vert
\psi\rangle$ with the echo operator $U^{t}:=U_{\epsilon}^{-t}U_{0}^{t}$, it
can be interpreted as the Loschmidt echo, i.e., an overlap of an initial state
with a state evolved for time $t$ with $H_{0}$ and subsequently for time $-t$
with $H_{\epsilon}$. (In general, we write time $t$ as a superscript.
Subscript $\epsilon$ denotes that $H_{\epsilon}$ was used for dynamics. If an
evolution operator, phase space coordinate, or density lacks a subscript
$\epsilon$, Loschmidt echo dynamics is implied.) QF amplitude~(\ref{fQM}) is
ubiquitous in applications: it appears in NMR spin echo experiments
\cite{nmr_echo_4}, neutron scattering~\cite{petitjean:2007}, ultrafast
electronic spectroscopy
\cite{mukamel:1982,*rost:1995,*li:1996,*egorov:1998,*shi:2005,wehrle:2011},
etc. QF (\ref{QF}) is relevant in QM computation and decoherence
\cite{cucchietti:2003,*gorin:2004}, and can be used to measure nonadiabaticity
\cite{zimmermann:2010a,*zimmermann:2011} or accuracy of molecular QD on an
approximate potential energy surface \cite{li:2009,*zimmermann:2010c}.

Definition (\ref{QF}) can be generalized to mixed states in different ways
\cite{gorin:2006,vanicek:2004a,vanicek:2004b,vanicek:2006}, but we assume that
the initial states are pure. In this case, one may write QF (\ref{QF}) as
$F_{\text{QM}}(t)=\operatorname{Tr}\left(  \hat{\rho}_{\epsilon}^{t}\hat{\rho
}_{0}^{t}\right)  $ where $\hat{\rho}_{\epsilon}^{t}:=U_{\epsilon}^{t}%
\hat{\rho}U_{\epsilon}^{-t}$ is the density operator at time $t$. In the
phase-space formulation of QM mechanics, QF becomes {$F_{\text{QM}}%
(t)={h}^{-{D}}\int dx\rho_{\epsilon,{\text{W}}}^{t}(x)\rho_{0,{\text{W}}}%
^{t}(x)$} where $x:=\left(  q,p\right)  $ is a point in phase space and
{{$A_{\text{W}}(x):=\int d\xi\langle q-\xi/2\left\vert \hat{A}\right\vert
q+\xi/2\rangle e^{ip\xi/\hbar}$}} is the Wigner transform of $\hat{A}$. This
form of QF suggests its CL limit, called CL fidelity (CF) \cite{prosen:2002a,
benenti1}
\begin{align}
F_{\text{CL}}(t) &  :=F_{\text{fid}}(t)=h^{-D}\int dx\rho_{\epsilon}%
^{t}(x)\rho_{0}^{t}(x)\label{CF_fid}\\
&  =F_{\text{echo}}(t)={h}^{-{D}}\int d{x}\rho^{t}(x)\rho^{0}%
(x)\label{CF_echo}%
\end{align}
where the first and second line express CF in the fidelity and Loschmidt echo
pictures, respectively. If $F$ or $\rho$ lack the subscript \textquotedblleft
CL\textquotedblright, \textquotedblleft QM\textquotedblright, or
\textquotedblleft DR\textquotedblright, \textquotedblleft CL\textquotedblright%
\ is implied.

\emph{Dephasing representation. }There were several attempts at describing QF
semiclassically. Most were analytical
\cite{jalabert:2001,*cohen_kottos:2000,*cerruti:2002,jacquod:2009} and valid
only under special circumstances because the numerical approaches were
overwhelmed with the sign problem. Extending a numerical SC method for
localized Gaussian wavepackets (GWPs) \cite{vanicek:2003a}, the DR was
introduced as a more accurate and general approximation of QF
\cite{vanicek:2004a,vanicek:2004b,vanicek:2006}. The DR of QF amplitude is an
interference integral
\begin{align}
f_{\text{DR}}(t) &  :=h^{-D}\int dx^{0}\rho_{\text{W}}(x^{0})\exp[i\phi
(x^{0},t)],\label{fDR}\\
\phi(x^{0},t) &  :=-\Delta S(x^{0},t)/\hbar=\left(  \epsilon/\hbar\right)
\int_{0}^{t}d\tau V(x_{\epsilon/2}^{\tau}),\label{DeltaS}%
\end{align}
where the phase $\phi$ is determined by the action $\Delta S$ due to the
perturbation along a trajectory propagated with the average Hamiltonian
$H_{\epsilon/2}$ \cite{wehrle:2011,zambrano:2011}. Above, $x_{\epsilon}%
^{t}:=\Phi_{\epsilon}^{t}(x^{0})$ where $\Phi_{\epsilon}^{t}$ is the
Hamiltonian flow of $H_{\epsilon}$ and the perturbation $V$ can, in general,
depend on both $q$ and $p$. The DR of fidelity, computed as $F_{\text{DR}%
}:=|f_{\text{DR}}|^{2}$, was successfully used to describe stability of QD in
integrable, mixed, and chaotic systems
\cite{vanicek:2004a,vanicek:2004b,vanicek:2006}, nonadiabaticity
\cite{zimmermann:2010a,zimmermann:2011} and accuracy of molecular QD on an
approximate potential energy surface \cite{li:2009,zimmermann:2010c}, and the
local density of states and the transition from the Fermi-Golden-Rule
(FGR)\ to the Lyapunov regime of QF decay
\cite{wang:2005,*ares:2009,*wisniacki:2010,*garcia-mata:2011b}. The same
approximation was independently derived and used in electronic spectroscopy
\cite{mukamel:1982,*rost:1995,*li:1996,*egorov:1998,*shi:2005}. Recently, the
range of validity of the DR was extended with a SC prefactor
\cite{zambrano:2011}. The remarkable efficiency of the original DR observed
empirically in applications led us to analyze this property rigorously here
and to compare it with the efficiencies of the QM and CL calculations of QF.

\emph{Algorithms. }The most general and straightforward way to evaluate
Eqs.~(\ref{CF_fid})-(\ref{CF_echo}) and (\ref{fDR}) is with trajectory-based
methods. While the DR (\ref{fDR}) is already in a suitable form,
Eqs.~(\ref{CF_fid})-(\ref{CF_echo}) for CF must be rewritten using the
Liouville theorem as
\begin{align}
F_{\text{fid}}(t)  &  ={h}^{-{D}}\int d{x^{0}}\rho(x_{\epsilon}^{-t}%
)\rho(x_{0}^{-t})\text{ \ and}\label{CF_fid_tr}\\
F_{\text{echo}}(t)  &  ={h}^{-{D}}\int d{x^{0}}\rho(x^{-t})\rho(x^{0}).
\label{CF_echo_tr}%
\end{align}
Above, $x^{t}:=\Phi^{t}(x^{0})$ where $\Phi^{t}:=\Phi_{\epsilon}^{-t}\circ
\Phi_{0}^{t}$ is the Loschmidt echo flow. Since it is the phase space points
rather than the densities that evolve in expressions (\ref{CF_fid_tr}%
)-(\ref{CF_echo_tr}), we can take $\rho=\rho_{\text{W}}^{t=0}$. For numerical
computations, Eqs.~(\ref{fDR}) and (\ref{CF_fid_tr})-(\ref{CF_echo_tr}) are
further rewritten in a form suitable for Monte Carlo evaluation, i.e., as an
average
\[
\left\langle A(x^{0},t)\right\rangle _{W(x^{0})}:=\frac{\int d{x^{0}}%
A(x^{0},t)W(x^{0})}{\int d{x^{0}}W(x^{0})}%
\]
where $W$ is the sampling weight for initial conditions $x^{0}$. Using
$W=\rho_{\text{W}}(x^{0})$, the DR algorithm becomes
\cite{vanicek:2004a,vanicek:2004b,vanicek:2006}
\begin{equation}
f_{\text{DR}}(t)={{\left\langle \exp\left[  i\phi(x^{0},t)\right]
\right\rangle }_{\rho_{\text{W}}(x^{0})}}. \label{FDR-av}%
\end{equation}
Sampling is straightforward for $\rho_{\text{W}}\geq0$, but can be done also
for general pure states \cite{vanicek:2006}. While previously used CL
algorithms sampled from $W=\rho$
\cite{benenti1,Karkuszewski:2002,*benenti:2003,*benenti:2003a,*veble:2004,*casati:2005,*veble:2005}%
, Ref. \cite{mollica:2011} considered more general weights $W=W_{M}%
(x^{0}):=\rho(x^{0})^{M}$ and $W=W_{M}(x_{0}^{-t})=\rho(\Phi_{0}^{-t}%
(x^{0}))^{M}$ for the echo and fidelity dynamics, respectively. These weights
yield four families of $M$-dependent algorithms \cite{mollica:2011}%
\begin{align}
F_{\text{fid-}M}(t)  &  =I_{M}{\langle\rho(x_{\epsilon}^{-t})\rho(}x_{0}%
^{-t})^{1-M}{\rangle}_{\rho(x_{0}^{-t})^{M}},\label{CF_fid_M}\\
F_{\text{echo-}M}(t)  &  =I_{M}{\langle\rho(x^{-t})\rho(}x^{0})^{1-M}{\rangle
}_{\rho(x^{0})^{M}},\label{CF_echo_M}\\
F_{\text{fid-N-}M}(t)  &  =\frac{{\langle\rho(x_{\epsilon}^{-t})\rho(}%
x_{0}^{-t})^{1-M}{\rangle}_{\rho(x_{0}^{-t})^{M}}}{{\langle}\rho(x_{0}%
^{-t})^{2-M}{\rangle}_{\rho(x_{0}^{-t})^{M}}},\label{CF_fid-N_M}\\
F_{\text{echo-N-}M}(t)  &  =\frac{{\langle\rho(x^{-t})\rho(}x^{0}%
)^{1-M}{\rangle}_{\rho(x^{0})^{M}}}{{\langle\rho(}x^{0})^{2-M}{\rangle}%
_{\rho(x^{0})^{M}}}, \label{CF_echo-N_M}%
\end{align}
where $I_{M}:=h^{-D}\int\rho(x^{0})^{M}dx^{0}$ is a normalization factor.
Conveniently, the \textquotedblleft normalized\textquotedblright\ (N)
algorithms (\ref{CF_fid-N_M})-(\ref{CF_echo-N_M}) do not require the
normalization factor $I_{M}$ which is, for general states, known explicitly
only for $M\in\{0,1,2\}$ ($I_{0}=n_{1}^{D}$, $I_{1}=I_{2}=1$). For further
details, see Ref. \cite{mollica:2011} where it was found that the echo-2
algorithm is optimal since it is already normalized (i.e., echo-2 = echo-N-2),
applies to any pure state (in particular, $\rho$ does not have to be positive
definite), and--most importantly--is by far the most efficient CL algorithm.

\emph{Efficiency. }The reader does not have to be persuaded of the exponential
scaling of QD with $D$. We just note that the direct diagonalization of the
Hamiltonian leads to a QD algorithm with a cost $O(t^{0}n_{D}^{3}%
)=O(t^{0}n_{1}^{3D})$ where $n_{D}=n_{1}^{D}$ is the dimension of the Hilbert
space of $D$ DOF. Despite the independence of $t$, the scaling with $D$ is
overwhelming. More practical is the split-operator algorithm requiring the
fast Fourier transform~(FFT) at each step. The complexity of FFT is
$O(n_{D}\log n_{D})$, hence the overall cost is $O(tD{n_{1}}^{D}\log{n_{1}})$.
The effective $n_{1}$ is reduced in increasingly popular methods with evolving
bases, but the exponential scaling remains.

Regarding the CF and DR algorithms, efficiency of trajectory-based methods
depends on two ingredients: First, what is the cost of propagating $N$
trajectories for time $t$? Second, what $N$ is needed to converge the result
to within a desired discretization error $\sigma_{\text{discr}}$? As this
analysis was done for the CL algorithms in Ref. \cite{mollica:2011}, here we
only outline the main ideas and apply them to analyze the efficiency of the DR.

The cost of a typical method propagating $N$ trajectories for time $t$ is
$O(c_{\text{f}}tN)$ where $c_{\text{f}}$ is the cost of a single force
evaluation. However, among the above mentioned algorithms, this is only true
for the fidelity algorithms with $M=0$ (i.e., fid-0 and fid-N-0) and for the
DR! Remarkably, in all other cases, the cost is $O(c_{\text{f}}t^{2}N)$. The
cost is linear in time for a single time $t$, but becomes quadratic if one
wants to know CF for all times up to $t$. For the echo algorithms, it is due
to the necessary full backward propagation for each time between $0$ and $t$.
For the fidelity algorithms, it is because the weight function $\rho
(x^{-t})^{M}$ changes with time and the sampling has to be redone for each
time between $0$ and $t$ \cite{mollica:2011}.

\begin{figure}
[hptb]\centerline{\resizebox{\hsize}{!}{\includegraphics[]{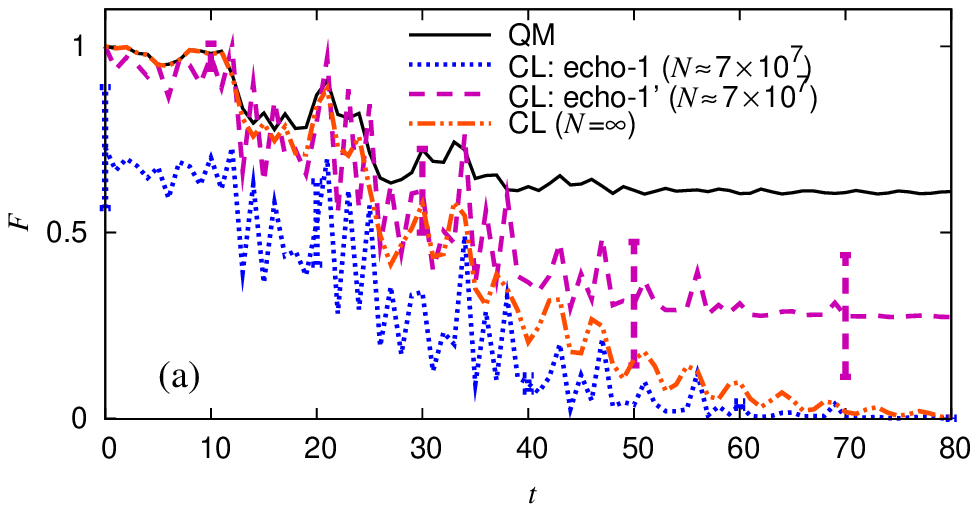}}}
\centerline{\resizebox{\hsize}{!}{\includegraphics[]{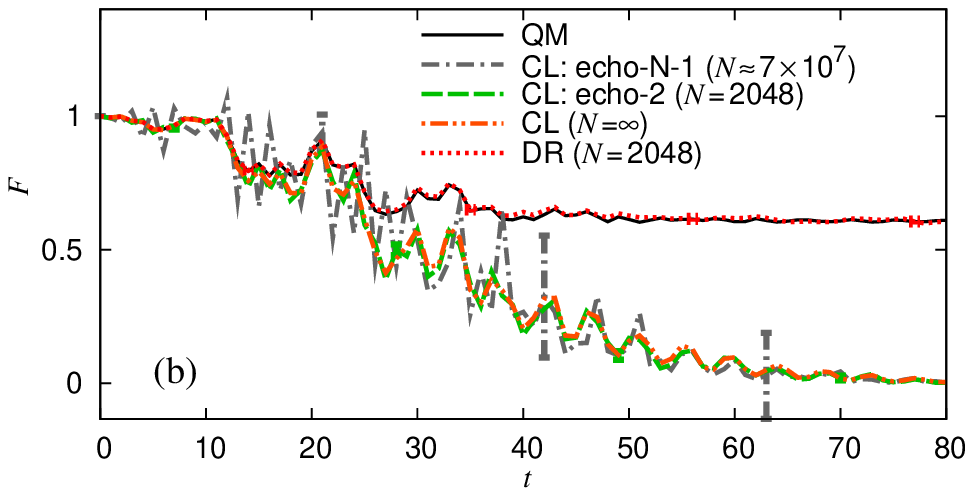}}}
\caption{\label{FIG1}Convergence of different fidelity algorithms in a $100$-dimensional system of
perturbed~($\epsilon= 3\times 10^{-4}$)
quasi-integrable~($k=0.2$) kicked rotors with $n_{1}=8192$.
Error bars plotted every 20 time steps. (a) Simple algorithms echo-1 and echo-1' are far from converged even
with $7 \times 10^7$ trajectories. (b) While both DR and echo-2 algorithms converge fully with only $2048$ trajectories,
only the DR can capture the QM fidelity ``freeze'' (the plateau).}
\end{figure}

\begin{figure}
[hptb]\centerline{\resizebox{\hsize}{!}{\includegraphics[]{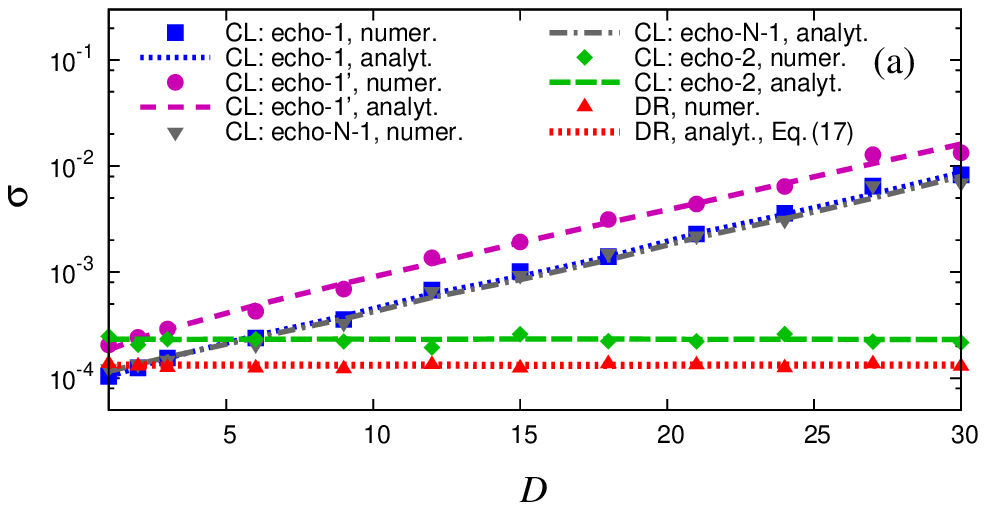}}}
\centerline{\resizebox{\hsize}{!}{\includegraphics[]{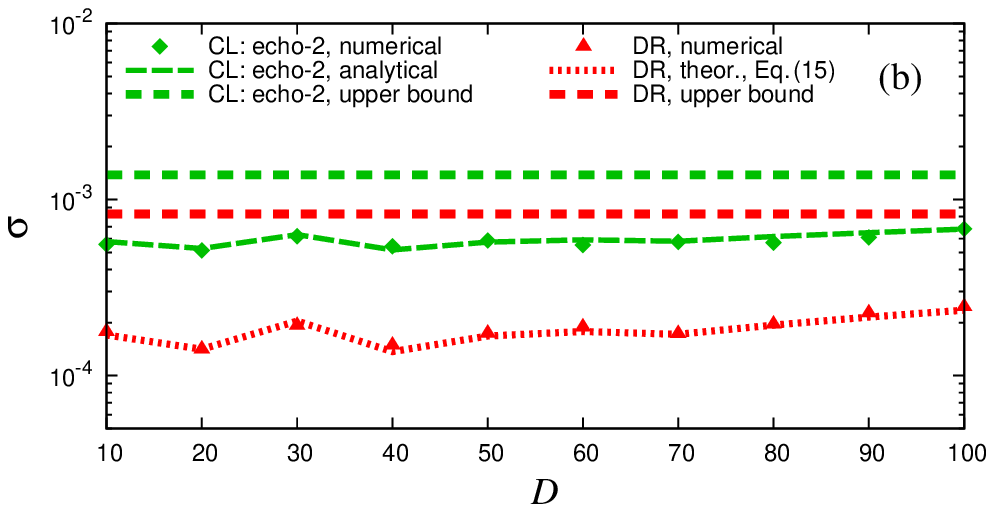}}}
\caption{\label{FIG2A}Statistical error grows exponentially with $D$ for the echo-1, echo-1', and echo-N-1
algorithms, while it is independent of $D$ for the echo-2 and DR algorithms. (a) Pure displacement dynamics
obtained with two displaced $D$-dimensional SHOs. $N \approx 10^7$. Time chosen separately for each $D$ so that $F\approx0.3$.
(b) General dynamics obtained with a $D$-dimensional system of perturbed ($\epsilon= 10^{-4}$) quasi-integrable~($k=0.2$)
kicked rotors with $n_{1}=131072$. $N \approx 5\times 10^5$. Time chosen separately for each $D$ so that $F\approx0.9$.}
\end{figure}

\begin{figure}
[hptb]\centerline{\resizebox{\hsize}{!}{\includegraphics[]{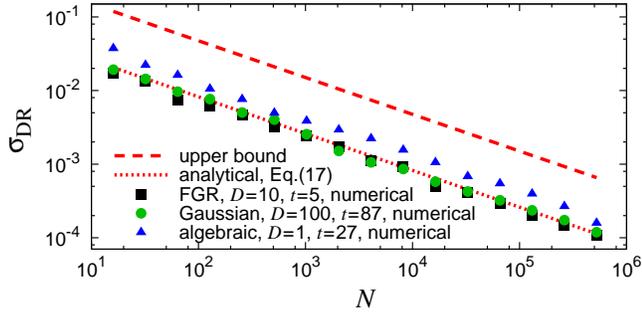}}}
\caption{\label{all-regimes} Regardless of dynamics, statistical error of the DR is independent of dimensionality ($D$), time
($t$), and is proportional to $N^{-1/2}$. Errors are compared
for $10$ kicked rotors in the chaotic FGR regime $(k=18,\epsilon
=6.4\times10^{-6},n_{1}=131072)$,  $100$ kicked rotors
in the integrable Gaussian regime $(k=0.2,\epsilon=6.4\times10^{-6}
,n_{1}=131072)$, and a single kicked rotor in the quasi-integrable algebraic
regime $(k=0.2,\epsilon=6.4\times10^{-4},n_{1}=131072)$. Time $t$ was
chosen separately for each system so that $F\approx0.94$.}
\end{figure}

\begin{figure}
[hptb]\centerline{\resizebox{\hsize}{!}{\includegraphics[]{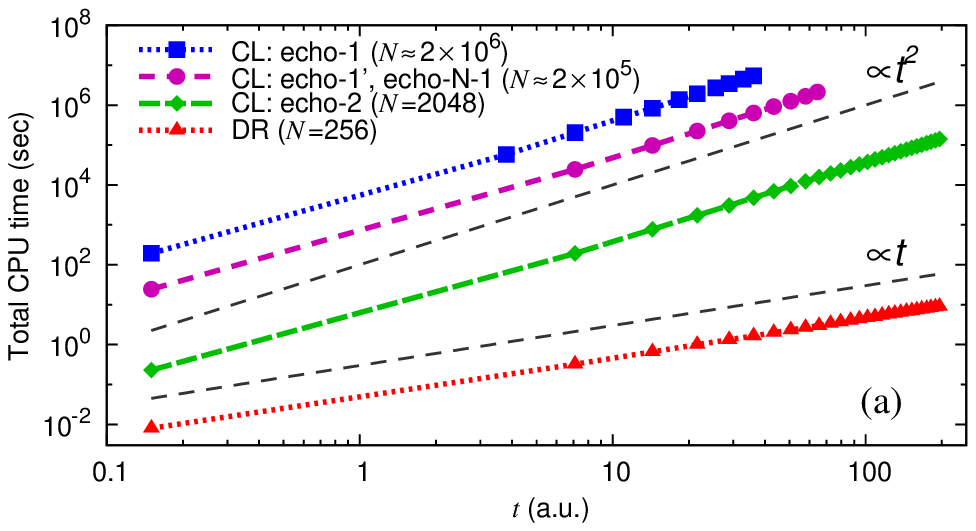}}}
\centerline{\resizebox{\hsize}{!}{\includegraphics[]{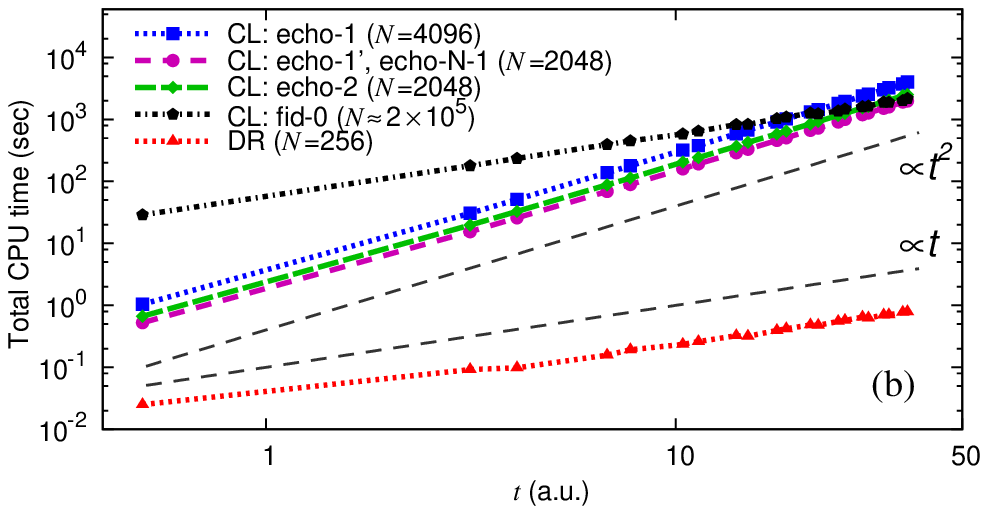}}}
\caption{\label{CPU-costs}Computational cost as a function of simulation time $t$ for different algorithms.
The CPU time grows quadratically with $t$ for all CL echo algorithms
while it is only linear with $t$ for the DR and fid-0 algorithms. Dynamics is of pure displacement type given
by a $D$-dimensional system of displaced SHOs for which all algorithms converge to the exact result.
Fidelity was computed at fidelity revival times at which $F\approx 0.9$. Number of
trajectories ($N$) was selected for each algorithm separately so that the statistical error $\sigma\approx0.01$
for all methods. (a) $D=20$. (b) $D=1$.}
\end{figure}

The number $N$ of trajectories required for convergence can depend on $D$,
$t$, dynamics, initial state, and method. Below we estimate $N$ for the DR
analytically using the technique proposed in Ref. \cite{mollica:2011}. The
expected systematic component of $\sigma_{\text{discr}}$ is zero for
$f_{\text{DR}}$ and $O(N^{-1})$ for $F_{\text{DR}}$ and is negligible to the
expected statistical component $\sigma=O(N^{-1/2})$ which therefore determines
convergence.\ Expected statistical error of $A(N)$ is computed as $\sigma
_{A}^{2}(N)={\overline{\left\vert A(N)\right\vert ^{2}}-}\left\vert
{\overline{A(N)}}\right\vert {^{2}}$ where the overline denotes an average
over infinitely many independent simulations with $N$ trajectories.

The discretized form of Eq.~(\ref{FDR-av}) is $f_{\text{DR}}(t,N)=N^{-1}%
\sum_{j=1}^{N}\exp[i\phi(x_{j}^{0},t)]$, from which $\overline{\left\vert
f_{\text{DR}}(t,N)\right\vert ^{2}}=N^{-1}+(1-N^{-1})F_{\text{DR}}(t)$,
$\left\vert \overline{f_{\text{DR}}(t,N)}\right\vert ^{2}=F_{\text{DR}}(t)$,
and $\sigma_{f_{\text{DR}}}^{2}=N^{-1}(1-F_{\text{DR}})$. The analogous
calculation for $F_{\text{DR}}$ is somewhat more involved but straightforward.
Inverting the results for $\sigma_{f_{\text{DR}}}^{2}$(exact) and
$\sigma_{F_{\text{DR}}}^{2}$ (to leading order in $N$) gives%
\begin{align}
N_{f_{\text{DR}}}  &  =\sigma^{-2}(1-F_{\text{DR}})\text{ and}%
\label{stat-fDR-analyt}\\
{N_{F_{\text{DR}}}}  &  ={\frac{2}{\sigma^{2}}\left[  \operatorname{Re}\left(
{\langle e^{i2\phi}\rangle_{\rho_{\text{W}}}{\langle e^{-i\phi}\rangle}%
_{\rho_{\text{W}}}^{2}}\right)  {+F_{\text{DR}}-2F_{\text{DR}}^{2}}\right]  .}
\label{stat-DR-theor}%
\end{align}
Result (\ref{stat-fDR-analyt}) for $N_{f_{\text{DR}}}$ is completely general.
As for $N_{F_{\text{DR}}}$, using the inequality $\left\vert \langle
e^{i2\phi}\rangle_{\rho_{\text{W}}(x^{0})}\right\vert \leq1$ and Eq.
(\ref{FDR-av}), we can find a completely general upper bound,
\begin{equation}
N_{F_{\text{DR}}}\leq4\sigma^{-2}F_{\text{DR}}(1-F_{\text{DR}}).
\label{stat-DR-bound}%
\end{equation}
Estimate (\ref{stat-fDR-analyt}) and upper bound (\ref{stat-DR-bound}) show,
remarkably, that without any assumptions, the numbers of trajectories needed
for convergence of both $f_{\text{DR}}$ and $F_{\text{DR}}$ depend only on
$\sigma$ and $F_{\text{DR}}$, and are\emph{\ independent} of $D$, $t$, initial
state, or dynamics. Estimate (\ref{stat-DR-theor}) of $N_{F_{\text{DR}}}$ can
be evaluated analytically for normally distributed phase $\phi$. This is
satisfied very accurately in the chaotic FGR and integrable Gaussian regimes
\cite{gorin:2006,jacquod:2009}, and exactly for pure displacement dynamics of
GWPs. Noting that {for normal distributions $\langle e^{i\phi}\rangle
=e^{i\langle\phi\rangle}{\exp[-\operatorname{Var}(\phi)/2]}$ and
$F_{\text{DR}}={|f_{\text{DR}}|}^{2}=\exp[-\operatorname{Var}(\phi)]$,}
Eq.~(\ref{stat-DR-theor}) reduces to
\begin{equation}
N{_{F_{\text{DR}}\text{,normal}}=2\sigma}^{-2}{F_{\text{DR}}}\left(
{{1-F_{\text{DR}}}}\right)  {^{2},} \label{stat-DR-analyt-normal}%
\end{equation}
which is again \emph{independent} of $D$, $t$, initial state, or dynamics.

Using a similar analysis, in Ref. \cite{mollica:2011} it was found that for
CF\ algorithms (\ref{CF_fid_M})-(\ref{CF_echo-N_M}) and $D\gg1$, one needs
${N=\sigma^{-2}\alpha(F}){\beta^{D}}$ trajectories where $\alpha$ and $\beta
${\ depend on the method, initial state, and dynamics. }For all methods with
$M\neq2$, there are simple examples \cite{mollica:2011} with $\beta>1$,
implying an exponential growth of $N$ with $D$. Remarkably, for any dynamics
and any initial state, $\beta=1$ for the echo-2 algorithm, implying, as for
the DR, that $N$ is independent of $D$ \cite{mollica:2011}.

\emph{Numerical results and conclusion. }To illustrate the analytical results,
numerical tests were performed in multidimensional systems of uncoupled
displaced simple harmonic oscillators (SHOs, for pure displacement\ dynamics)
and perturbed kicked rotors (for nonlinear integrable and chaotic dynamics).
The last model is defined, $\operatorname{mod}${$(2\pi)$}, by the map
{$q_{j+1}=~q_{j}+p_{j}$}, {$p_{j+1}=~p_{j}-\nabla W(q_{j+1})-\epsilon\nabla
V(q_{j+1})$} where {$W(q)=-k\cos q$} is the potential and {$V(q)=-\cos(2q)$}
the perturbation of the system; $k$ and $\epsilon$ determine the type of
dynamics and perturbation strength, respectively. Uncoupled systems were used
in order to make QF calculations feasible (as a product of $D$ 1-dimensional
calculations); however, both CF and DR calculations were performed as for a
truly $D$-dimensional system. The initial state was always a multidimensional
GWP. Expected statistical errors were estimated by averaging actual
statistical errors over $100$ different sets of $N$ trajectories. No fitting
was used in any of the figures, yet all numerical results agree with the
analytical estimates. Note that the figures show also results for algorithm
echo-1', $F_{\text{echo-1'}}(t)=1+{\langle\rho(x^{-t})-\rho(x^{0})\rangle
}_{\rho(x^{0})},$ which is a variant of echo-1 accurate for high fidelity
~\cite{mollica:2011}.

Figure~\ref{FIG1} displays fidelity in a $100$-dimensional system of kicked
rotors. It shows that both echo-2 and DR algorithms converge with several
orders of magnitude fewer trajectories than the echo-1, echo-1', and echo-N-1
algorithms but while the DR agrees with the QM result, even the fully
converged CF (computed as a product of 100 one-dimensional fidelities) cannot
reproduce QM effects. Figure~\ref{FIG2A} confirms that whereas the statistical
errors of the echo-1, echo-1', and echo-N-1 algorithms grow exponentially with
$D$, statistical errors of the DR and echo-2 algorithms are independent of
$D$. Figure~\ref{all-regimes} shows that for several very different dynamical
regimes, $\sigma_{\text{DR}}$ is independent of $t$, $D$ and $n_{1}$, in
agreement with the general upper bound (\ref{stat-DR-bound}) and--in the FGR
and Gaussian regimes--also in agreement with the analytical estimate
(\ref{stat-DR-analyt-normal}). Finally, figure~\ref{CPU-costs} exhibits the
superior computational efficiency of the DR compared to all CF algorithms:
thanks to the linear scaling with $t$ and independence of $D$, the DR is
orders of magnitude faster already for quite a small system and short time.

To conclude, in the case of QF, a SC method can be not only more accurate, but
also more efficient than a CL simulation of QD. This counterintuitive result
should be useful for future development of approximate methods for QD of large
systems. This research was supported by Swiss NSF grant No. 200021\_124936 and
NCCR MUST, and by EPFL. We thank T. Seligman and T. Zimmermann for discussions.

\bibliographystyle{apsrev4-1}

\end{document}